\documentclass[%
 superscriptaddress,twocolumn,10pt,prx,aps,showpacs,longbibliography,floatfix]{revtex4-2}

\usepackage{soul}
\usepackage{almpaper}
\usepackage{siunitx}
\usepackage{booktabs}

\newacronym{dfs}{DFS}{decoherence-free subspace}
\newacronym{bqc}{BQC}{bosonic quantum code}

\graphicspath{{paperfigs/}}

\begin{document}


\title{Chiral cat code: Enhanced error correction induced by higher-order nonlinearities}


\author{Adrià Labay-Mora}
\email{alabay@ifisc.uib-csic.es}
\affiliation{Institute for Cross-Disciplinary Physics and Complex Systems (IFISC) UIB-CSIC, Campus Universitat Illes Balears, 07122 Palma de Mallorca, Spain}
\author{Alberto Mercurio}
\affiliation{Institute of Physics, Ecole Polytechnique Fédérale de Lausanne (EPFL), CH-1015 Lausanne, Switzerland}
\affiliation{Center for Quantum Science and Engineering, Ecole Polytechnique Fédérale de Lausanne (EPFL), CH-1015 Lausanne, Switzerland}
\author{Vincenzo Savona}
\affiliation{Institute of Physics, Ecole Polytechnique Fédérale de Lausanne (EPFL), CH-1015 Lausanne, Switzerland}
\affiliation{Center for Quantum Science and Engineering, Ecole Polytechnique Fédérale de Lausanne (EPFL), CH-1015 Lausanne, Switzerland}
\author{Gian Luca Giorgi}
\affiliation{Institute for Cross-Disciplinary Physics and Complex Systems (IFISC) UIB-CSIC, Campus Universitat Illes Balears, 07122 Palma de Mallorca, Spain}%
\author{Fabrizio Minganti}
\email{fabrizio.minganti@gmail.com}
\affiliation{Institute of Physics, Ecole Polytechnique Fédérale de Lausanne (EPFL), CH-1015 Lausanne, Switzerland}
\affiliation{Alice \& Bob,  53 Boulevard du G\'en\'eral Martial Valin, 75015, Paris, France}

\date{\today}
\begin{abstract}
We introduce a chiral Schr\"odinger cat qubit, a novel bosonic quantum code generalizing Kerr cat qubits that exploits higher-order nonlinearities.
Compared to standard Kerr cats, the chiral cat qubit allows additional correction of bit-flip errors within the Hilbert space of a single bosonic oscillator.
This property results from optical bistability, i.e., the simultaneous presence of multiple long-lived states. 
Two of them define the code logical basis and two define an error manifold. 
Thanks to the chiral structure of the phase space of this system, the error manifold can be engineered to ``capture'' bit flip events in the code space (a bit-flip trap), without affecting the quantum information stored in the system.
Therefore, it is possible to perform detection and correction of errors.
We demonstrate how this topological effect can be particularly efficient in the presence of large dephasing. 
We provide concrete examples of the performance of the code and show the possibility of applying quantum operations rapidly and efficiently.
Beyond the interest in this single technological application, our work demonstrates how the topology of phase space can enhance the performance of bosonic codes.
\end{abstract}

\maketitle

\section{Introduction}


Fault-tolerant quantum computing relies on the development of efficient error-correction protocols, creating logical qubits through redundant encoding of information \cite{nielsen2011,campbell2017,terhal2015}, protecting the system from the detrimental effect of the environment \cite{lidar2013, haroche2013, preskill2018, breuer2007}.
A mainstream approach achieves redundancy using multiple physical qubits (two-level systems).
Upon successful completion of an error correction protocol, 
the decay rate of the quantum information stored in the logical qubit is lower than that of the original components.
Superconducting circuits are candidates for the development of error-corrected quantum hardware \cite{Reed2012,ofek2016,Krinner2022,google2023,google2024,Putterman2025}. The two main mechanisms inducing errors are dephasing and photon loss noise, which combine to generate bit- and phase-flips.
Photon loss is due to the coupling of the system to spurious electromagnetic degrees of freedom (intrinsic losses) and coupling to various elements such as driving feedlines (extrinsic losses). Both induce loss of energy from the resonator. While proper isolation and filtering can reduce extrinsic losses, intrinsic ones can still severely limit the lifetimes of excitations in superconducting devices.
Dephasing, instead, mainly emerges when coupling the device to nonlinear elements \cite{noh2022}, fluctuation in gate charge in Josephson junctions, or as a consequence of magnetic-flux noise in flux-tunable configurations (e.g., the flux threading the SQUID loop) \cite{blais2021cqe}.
Given the rapid decay of quantum information even in state-of-the-art devices, error correction requires the fabrication and control of many physical qubits. 
The simultaneous correction of bit- and phase-flip errors was recently achieved by realizing a surface code, a highly controllable 2D geometry of qubits encoding the logical state in tens of data qubits and ancillas \cite{google2024}.
Ultimately, realizing logical operation between error-corrected qubits would require connecting and controlling many components, thus posing a major challenge towards the scaling-up of these devices.

Bosonic quantum codes (BQCs) offer a promising simplification of this complex challenge \cite{gottesman2001, mirrahimi2014,albert2018, cai2021, joshi2021, terhal2020, knill2000, michael2016}. Rather than storing information in the large Hilbert space of multiple physical qubits, BQCs achieve redundancy in the larger Hilbert spaces of a single bosonic oscillator.
A key example are Schr\"odinger cat qubits that use coherent states of opposite phase as logical codewords, with photonic Sch\"odinger cats representing the logical $x$ states. 
When successfully operated, cat qubits are noise biased as they efficiently suppress bit-flip errors while still being susceptible to phase-flip ones \cite{gilles1994,cochrane1999, mirrahimi2014}.
The size of the cat (i.e., the number of photons) is the key parameter determining the efficiency of the encoding. 
Larger cats are expected to exponentially reduce the rate of bit-flip errors at the price of a linear increase in the rate of phase flips.
A possible strategy then relies on the preparation of sufficiently large and controlled cat states, almost immune to bit-flip errors and requiring only the correction of phase flips \cite{guillaud2019,guillaud2021,guillaud2022}.
This task can be achieved using concatenated cats in a repetition code. By requiring fewer physical qubits per logical qubit, this architecture is expected to significantly reduce the hardware footprint with respect to a surface code, as recently experimentally demonstrated \cite{Putterman2025}. More advanced error correction schemes have also been recently proposed \cite{Ruiz2025}. However, coupling multiple nonlinear elements can lead to an increase in dephasing noise, as discussed for instance in Ref.~\cite{Putterman2025}.

In superconducting architectures operating cat states rely on an effective two-photon driving mechanism, together with two-photon dissipation (dissipative cat \cite{mirrahimi2014, leghtas2015, touzard2018, xu2022, goto2016}), Kerr nonlinearity (Kerr cat \cite{puri2017,puri2019, puri2020, grimm2020, frattini2022, Venkatraman2022}), or a combination of the two (hybrid cat \cite{gautier2022,ruiz2022,gravina2023criticalcat}) that stabilize the logical manifold spanned by coherent states.
This task requires nonlinear and parametric processes.
The former can be tailored by combining Josephson junctions, resulting in an anharmonic structure of the energy level of the resonator; within a first approximation, the potential of these nonlinear elements can be described as a photon-photon interaction term, known as Kerr nonlinearity.
Parametric processes result from applying external drives to nonlinear elements.
For example, two-photon pump and two-photon dissipation mechanisms can be generated by engineering a two-photon exchange between the memory, hosting the cat state, and a very dissipative mode (the buffer) \cite{leghtas2015,lescanne2020,Berdou2023}.

What differentiates Kerr, dissipative, and hybrid cats is their operations (gates) and response to errors, with dephasing and photon loss degrading the quantum information stored in the logical qubit.
Pioneering works on cat states, both Kerr and dissipative, demonstrated their possibilities and value for quantum information tasks \cite{leghtas2015,grimm2020}. 
However, Kerr cats are very susceptible to dephasing, as also 
discussed below.
More recently, several theoretical proposals have explored how additional features, such as squeezing \cite{schlegel2022,Hillmann2023,Xu2023,labay2023squeezed}, Hamiltonian terms \cite{gravina2023criticalcat,ruiz2022}, or engineered dissipation/bath engineering \cite{gautier2022,Gautier2023,rojkov2024stabilizationcatstatemanifoldsusing}, can provide additional protection from these dissipative events, with experimental results supporting these claims \cite{RousseauArXiv25}.
However, these models are often idealizations of actual devices, where the higher-order effects of the nonlinear elements are neglected.

\begin{figure}
    \centering
    \includegraphics[width=\linewidth]{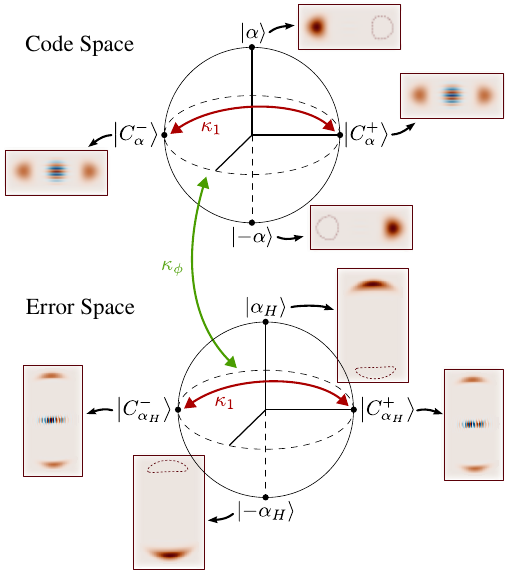}
    \caption{Bloch sphere representation of the code and error spaces of a chiral cat, spanned by the low- ($\ket{\pm \alpha}$) and high-amplitude ($\ket{\pm \alpha_H}$) lobes, respectively.
    Upon an appropriate choice of the system parameters, dephasing and photon loss induce a passage from the low- to the high-amplitude lobes (a bit-flip trap that captures the state without degrading quantum information), making it possible to perform error detection and correction.}
    \label{fig:bloch_chiral}
\end{figure}

In this article, we show that higher-order nonlinear processes can also be harnessed to improve the performance of the logical qubit, providing a pathway to protect Kerr cats from errors. 
Two facts that both depend on the simultaneous presence of lower- and higher-order nonlinearities permit this advantage:
(i) The competition between nonlinear terms can generate two metastable manifolds, characterized by a different phase and photon number, and in each of them quantum information can be stored; 
(ii) Within an appropriate choice of parameters, the phase space of the bosonic system can be endowed with a chiral structure.
Thus, the emergent structure is one where two quasiorthogonal code and error spaces coexist (see Fig.~\ref{fig:bloch_chiral}), with the error space capturing bit-flip errors.
As we show below, the presence of these features permits an error correction of bit-flip errors, significantly reducing the error rates of Kerr cats, even in the presence of large dephasing.
Our work shows that states with Kerr and higher-order nonlinearities can still be used as logical qubits, and even outperform idealized Kerr cats.
All the numerical simulations of the full quantum model were carried out using the \texttt{QuantumToolbox.jl} package in Julia \cite{alberto_mercurio_2024_quantumtoolbox}.

\section{Model and working principle}

The system we consider is a nonlinear bosonic resonator, driven by a two photon pump of amplitude $\epsilon_s$, which in the frame of the pump reads (setting $\hbar = 1$)
\begin{equation}\label{Eq:Hamiltonian}
    \hmt = -\Delta \hat{a}^\dagger \hat{a} + \frac{K_2}{2} \, \,(\hat{a}^\dagger)^2 \hat{a}^2 + \frac{K_3}{3} \, \, (\hat{a}^\dagger)^3 \hat{a}^3 + \epsilon_s \, [(\hat{a}^\dagger)^2 + \hat{a}^2 ].
\end{equation}
Here, $\Delta$ is the pump-to-cavity detuning, $K_2 $ is the Kerr term describing boson-boson interaction, and $K_3 $ is the first higher-order correction describing three-body processes. 
Higher-order nonlinearities, of the form $K_n (\hat{a}^\dagger)^{n} a^{n}$, can also be included.
This Hamiltonian typically emerges when developing the sinusoidal potential of nonlinear superconducting elements based on Josephson junctions, such as SQUIDs and SNAILs \cite{Blais2021}.
Normally, $K_2$ dominates over other nonlinearities, which are in many cases neglected.

\begin{figure*}[t]
    \includegraphics[width =\linewidth]{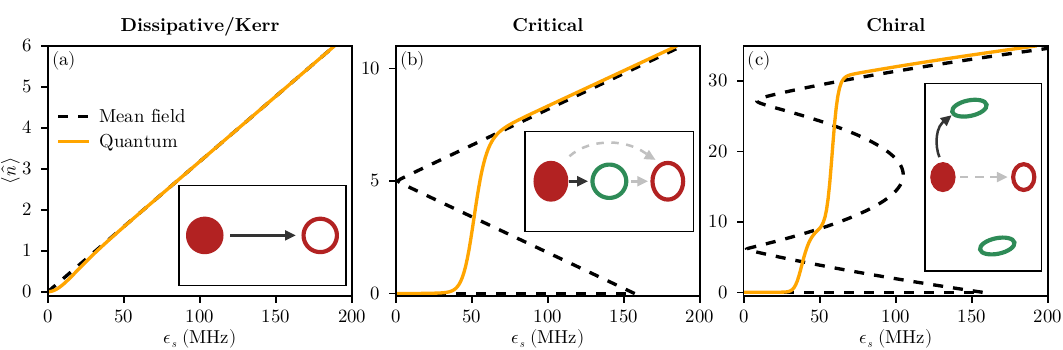}
    \caption{Schematics of various types of cat states. The panels show the mean photon number as a function of the two-photon drive amplitude. Solid lines represent the steady-state photon number, while the dashed line represents the prediction of the semiclassical approximation. The insets show a sketch of the Wigner function, with the solid arrows indicating main the error processes and dashed ones depicting suppressed errors.
    (a) For a pure dissipative~\cite{leghtas2015} Kerr~\cite{grimm2020} cat, increasing the two-photon drive linearly increases the photon number. For a given drive, the system spans a manifold characterized by two states of opposite phase $\rho_{\pm \alpha} \simeq \ketbra{\pm \alpha}$. Bit flip errors than take the form of a passage between these two states.
    (b) For a critical cat \cite{gravina2023criticalcat} (a hybrid cat is simultaneously stabilized by Kerr and two-photon loss and operated in the presence of detuning), the system displays optical bistability, i.e., the simultaneous presence of multiple solution according to the semiclassical approximation.
    The optimal regime of operation is one where a cat state will eventually decay into the vacuum, and quantum information is encoded in the metastable manifold spanned by squeezed states of opposite phases.
    The dominant source of errors is a leakage, corresponding to a jump to the vacuum ($\rho_{\pm\alpha} \to \ketbra{0}$).
    (c) The chiral cat is a hybrid cat with additional higher-order nonlinearity.
    The manifold $\rho_{\pm \alpha} \simeq \ketbra{\pm \alpha}$ coexists with a larger-photon number one at $\rho_{\pm \alpha_H} \simeq \ketbra{\pm \alpha_H}$, both capable of hosting cat-like states. The high manifold acts as a \textit{bit-flip trap} where the jump $\rho_{\alpha} \leftrightarrow \rho_{-\alpha}$ is suppressed in favour of $\rho_{\pm \alpha} \to \rho_{\pm \alpha_H}$. Finally, as the photon number between the two manifolds is different, the error can be detected and corrected. Parameters $(\mathrm{MHz})$: $\kappa_1/2\pi = 0.005$, $K_2/2\pi = -5$, $\kappa_\phi = 0$ and incrementally we set (a) $\Delta = K_3 = 0$, (b) $\Delta/2\pi = -25$ and $\kappa_2 / \abs{K_2} = 0.01$, and (c) $K_3/2\pi = 0.15$.}
    \label{fig:main-sketch}
\end{figure*}

The system is subject to the action of the environment, whose effect can be described by the Lindblad master equation:
\begin{equation}\label{eq:full_liouvillian}
    \mathcal{L} \rho = - i [\hmt, \rho] + \kappa_2 \mathcal{D}[\hat{a}^2] \rho + \kappa_1 \mathcal{D}[\hat{a}] \rho + \kappa_{\phi} \mathcal{D}[\hat{a}^\dagger \hat{a}] \rho.
\end{equation}
Here, $\kappa_1$ is the single photon loss rate, $\kappa_2$ is the rate of two-photon dissipation, and $\kappa_{\phi}$ is the dephasing rate.
While $\kappa_1$ and $\kappa_{\phi}$ tend to be determined by the design and fabrication of the device,  $\kappa_2$ can be obtained by reservoir engineering.
Typically, a second mode characterized by large photon loss rate, called the buffer, is used to tune two-photon dissipation \cite{leghtas2015}.
Here and in the following, we will consider a Kerr-dominated hybrid confinement, assuming $\abs{K_2} \gg \kappa_2>0$.
Furthermore, we will consider configurations with large $\kappa_{\phi}$, imagining concatenated architectures of highly-nonlinear and driven elements, where dephasing is expected to be further increased \cite{Putterman2025} with respect to that of a single component.

\subsection{Kerr, dissipative, and critical cats}\label{sec:kerr_dissipative_critical}

In this part, we briefly review the operating principles of ideal Kerr and dissipative cats by neglecting the effect of high-order nonlinearities ($K_3 = 0$). 

\subsubsection{Cats in the absence of photon loss and dephasing}

In this configuration, both dissipative, Kerr, and hybrid cats display similar properties, and the combined action of $\epsilon_s$, $K_2$, and $\kappa_2$ makes it possible to stabilize cat states if $\Delta = \kappa_1 = \kappa_{\phi} =0$. The even and odd cat states are defined by $\ket{C^{\pm}_{\alpha}} \propto \ket{\alpha} \pm \ket{-\alpha} $, with $\ket{\alpha} $ a coherent state such that $\hat{a}\ket{\alpha } = \alpha \ket{\alpha}$. Any combination of cat states is also stationary and the system admits two steady states of the form $\rho_{\rm ss} = \ketbra{C^{\pm}_{\alpha}}$ and two steady coherences $\rho_{\rm sc} = \ketbra{C^{\pm}_{\alpha}}{C^{\mp}_{\alpha}}$ with $\alpha = \sqrt{-\epsilon_s/(K_2 - i \kappa_2)}$ \cite{albert2016}.
Thus, the entire manifold spanned by cat (and coherent) states is stable, generating what is called a decoherence-free subspace \cite{albert2014,knill2000}.
Finally, in this configuration, but with $\Delta \neq 0$ and $\kappa_{\phi} \neq 0$ the system will still host two steady states, but no steady coherences.
In this idealized picture, we can define the ``standard'' representation of cat states on the Bloch sphere where we set the logical $\ket{\pm_z} = \ket{\pm \alpha}$, while $\ket{\pm_x} = \ket{C_\alpha^{\pm}}$ (see the definition of the code space in \cref{fig:bloch_chiral}).

Even if the cat properties as a memory do not change between dissipative and Kerr cats, different gates have to be tailored for each configuration.
Furthermore, the presence of additional Hamiltonian or dissipative terms can induce errors during the performance of logical operations.
For example, the $Z$ gate requires the action of a one-photon drive of the form $\epsilon_z (\hat{a} + \hat{a}^\dagger)$.
This term induces leakage outside of the cat manifold, whose magnitude depends on whether cats are generated by a Hamiltonian or a dissipative confinement.
In particular, one can show that Kerr nonlinearity makes the cat more resilient to $\epsilon_z$, allowing for stronger drives and thus faster gates than a dissipative cat \cite{gautier2022}.

\subsubsection{The effect of photon loss and dephasing}

This picture drastically changes in the presence of photon loss and dephasing that degrade the quantum information stored in the cat. 
A single steady state 
$\rho_{\rm ss} \simeq (\ketbra{C^{+}_{\alpha}} + \ketbra{C^{-}_{\alpha}})/2$ characterizes the system \cite{minganti2016,roberts2020}.
In the Bloch sphere, any point on the surface will decay towards the center, and dissipation will eventually lead to a completely mixed logical state.
We can understand this mixture as the simultaneous occurrence of bit-flip errors $\ket{\alpha} \to \ket{-\alpha}$ [sketched in Fig.~\ref{fig:main-sketch}(a)] and phase-flip errors $\ket{C^{\pm}_{\alpha}} \to \ket{C^{\mp}_{\alpha}}$. Here, $|\alpha|^2$ still grows linearly with the amplitude of the drive.
Cats are then predicated on the idea that, increasing their size, i.e., their photon number, bit-flip errors should be exponentially suppressed (coherent states are long-lived). In contrast, the rate of phase-flip errors increases linearly with the cat size.
The advantageous scaling of bit- over phase-flip errors makes cat states biased-noise qubits.

In all configurations, and even for the chiral cat introduced in the following, it can be shown that the rate of phase-flip errors scales as $\kappa_1 \expval*{\hat{a}^\dagger \hat{a}}$.
To determine the bit-flip rate, it is practical to resort to the Liouvillian eigenspectrum. Indeed, the Liouvillian is a linear superoperator, that can be diagonalized to retrieve the eigenvalues $\lambda_j$ and the eigenoperators $\rho_j$, defined by $\lv \rho_j = \lambda_j \rho_j$. The model is further characterized by a $Z_2$ weak symmetry (an invariance of the equation of motion \cite{albert2014}) as $\hat{a} \to -\hat{a}$ leaves Eq.~\eqref{eq:full_liouvillian} unchanged.
This, in turn, implies that, for each eigenoperator, $\hat{\Pi} \rho_j \hat{\Pi} = \pm \rho_j$, with $\hat{\Pi} = \exp(i \pi \hat{a}^\dagger \hat{a})$.
We can thus associate the quantum number $k = \pm 1$ to each eigenvalue and eigenoperator, to get
\begin{equation} \label{Eq:Eigenspectrum}
    \lv \rho_j^{(k)} = \lambda_j^{(k)} \rho_j^{(k)}, \quad k = \pm 1 .
\end{equation}
In particular, the Liouvillian eigenvalue $\lambda_0^{(-1)}$ describes the bit-flip error rate \cite{gravina2023criticalcat}, where we have ordered the eigenvalues by their real part [$|\Re(\lambda_0^{(k)})| = 0< |\Re(\lambda_1^{(k)})| < |\Re(\lambda_2^{(k)})| $].

\begin{figure}[t]
    \centering
    \includegraphics[width=\linewidth]{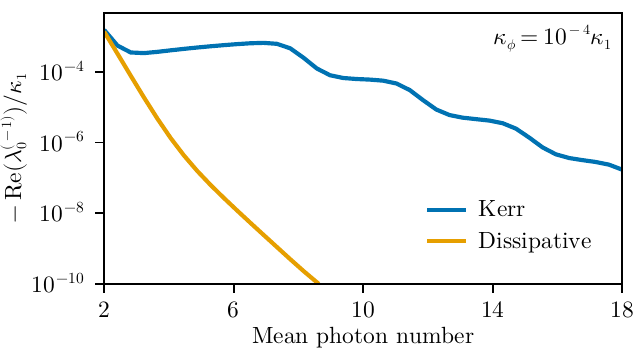}
    \caption{The Liouvillian gap describing the bit-flip error rate as a function of the mean photon number ($\abs{\alpha}^2$) for the dissipative (orange) and Kerr (blue) cats in a regime with negligible dephasing. Parameters as in \cref{fig:main-sketch} except $\epsilon_s$ which is chosen to match the mean photon number.}.
    \label{fig:Kerr_vs_dissipative_no_dephasing}
\end{figure}

First, we study the bit-flip error rates plotting $\lambda_0^{(-1)}$ in Fig.~\ref{fig:Kerr_vs_dissipative_no_dephasing} with moderate values of dephasing.
For dissipative cats, the bit-flip error rate becomes exponentially smaller with the number of photons.
For Kerr cats, despite an initial suppression of the bit-flip rate, increasing the photon number does not lead to the same exponential suppression of bit-flip errors as in the dissipative case.
Nevertheless, the Kerr cat shows protection from bit-flip errors with a ``staircase''-like behavior of $\lambda_0^{(-1)}$, observed and discussed in, e.g., \cite{Frattini2024}.

\begin{figure*}[t]
    \centering
    \includegraphics[width=\linewidth]{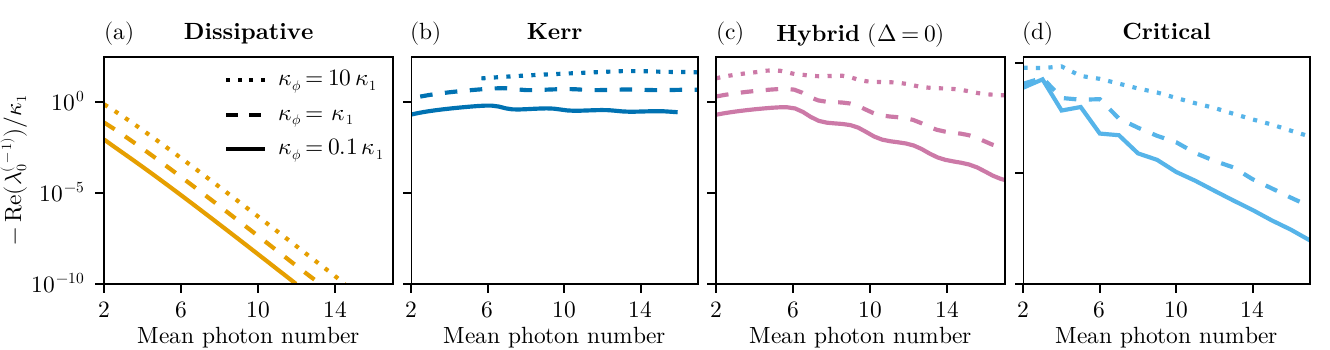}
    \caption{As a function of the mean-photon number, the Liouvillian gap describing the bit-flip error rate in (a) dissipative, (b) Kerr, (c) hybrid and (d) critical cat. We show three different regimes: single-photon dominated at $\kappa_\phi = 0.1 \kappa_1$ (solid lines), equal error rate $\kappa_\phi = \kappa_1$ (dashed lines) and dephasing dominated $\kappa_\phi = 10 \kappa_1$ (dotted lines).
    Parameters as in \cref{fig:main-sketch}. Detuning is zero for dissipative, Kerr and hybrid cats, but an optimal detuning is chosen for the critical cat.}
    \label{fig:Kerr_vs_dissipative}
\end{figure*}

The error suppression capability of cat states worsens in the presence of sizeable dephasing (see Fig.~\ref{fig:Kerr_vs_dissipative}).
For dissipative cats in Fig.~\ref{fig:Kerr_vs_dissipative}(a), the suppression of bit-flip errors is still exponential in the photon number, even if not as efficient as in the photon-loss dominated case.
Nonetheless, dissipative cats remain a promising platform for biased-noise qubit repetition codes, and candidates for low-hardware footprint error correction strategies.
For Kerr cats [Fig.~\ref{fig:Kerr_vs_dissipative}(b)], instead, the state loses its capability to suppress bit-flip errors, and thus it does not show all the beneficial properties that would make it a biased-noise qubit. 
This can be a major challenge towards the concatenation of multiple Kerr cats, as dephasing is expected to emerge in nonlinear coupled systems.

\subsubsection{Phase space structure}

A useful tool to understand errors and their effects is to use a combination of quantum trajectories (see details in the Appendix~\ref{App:Quantum_trajectories}) and semiclassical analysis.
Quantum trajectories, unlike master equation approaches that directly evolve the system's density matrix,  describe individual wave functions $\ket{\psi(t)}$ through sequences of deterministic and probabilistic events.
For example, consider a state initialized in $\ket{\alpha}$.
As discussed in the Appendix~\ref{App:Quantum_trajectories}, around this initial state, dissipation and Hamiltonian terms act as noise displacing an initial coherent state, with dephasing inducing large fluctuations in the form of temperature-like jumps, whose rate scales as $\kappa_{\phi} |\alpha|$.

To determine the effect of dephasing, we can then use a semiclassical approximation.
The semiclassical (coherent state) approximation neglects quantum fluctuations, assuming $\rho(t) = \ketbra{\alpha(t)}$ with $\ket{\alpha (t)}$ the coherent state. 
The evolution of the system, where we neglect dephasing $\kappa_\phi$ and high-order nonlinearity $K_3$, is captured by the C-number equation 
\begin{equation}
    \begin{split}
        \dot{\alpha}(t) =&  - \left(\frac{\kappa_1}{2} - \kappa_2 \abs{\alpha (t)}^2\right)\alpha (t) \\
        &- i 2 \epsilon_s \alpha^* (t) - i   K_2  \abs{\alpha(t)}^2 \alpha (t) 
    \end{split}
\end{equation}
We then study the vector field representing $ \dot{\alpha}(t)$, i.e., the direction of the evolution of a point in phase space.
Similar analysis has been used in, e.g., Refs.~\cite{villa2024topologicalclassificationdrivendissipativenonlinear, Cohen2023,ChvezCarlos2024,ferrari2023steadystatequantumchaosopen,ferrari_chaos_2024_chains} to study chaos and topological effects in driven two-photon systems.
We now have to compare the orbits of this system with the thermal-like noise induced by dephasing.

In a dissipative cat, shown in Fig.~\ref{fig:semiclassical}(a), the vector field will \textit{always} point back to one of the coherent states. For the cat to display a bit-flip error, large fluctuation are required, making the jump improbable.
The situation is drastically different in a Kerr cat, where the system forms periodic orbits around the minima of the potential.
This is shown in Fig.~\ref{fig:semiclassical}(b).
Small fluctuations can thus accumulate, eventually leading to bit-flip errors.

\subsubsection{Hybrid and critical cats}

Hybrid cats use both Kerr and dissipative confinement.
Considering a hybrid configuration where $\abs{K_2} \gg \kappa_2 > \kappa_{\phi}$, but with $\Delta =0$, makes the cat partially resilient to dephasing, suppressing bit flip errors by increasing the cat size \cite{gautier2022}.
This is shown in \cref{fig:Kerr_vs_dissipative}(c).
However, this configuration is still severely limited in its bit-flip error correction capability when compared to a fully dissipative cat.
Several works then noticed that detuning can be a resource to enhance the performance of a hybrid cat \cite{ruiz2022,Frattini2024,gravina2023criticalcat}.
Indeed, an appropriate choice of detuning suppresses bit-flip errors, as shown in \cref{fig:Kerr_vs_dissipative}(d).
This feature allows for accessing specific spectral degeneracies in cat states \cite{gautier2022,Frattini2024} and enables better error correction thanks to the introduction of squeezing \cite{schlegel2022}.

\begin{figure*}[t]
    \centering
    \includegraphics[width=\linewidth]{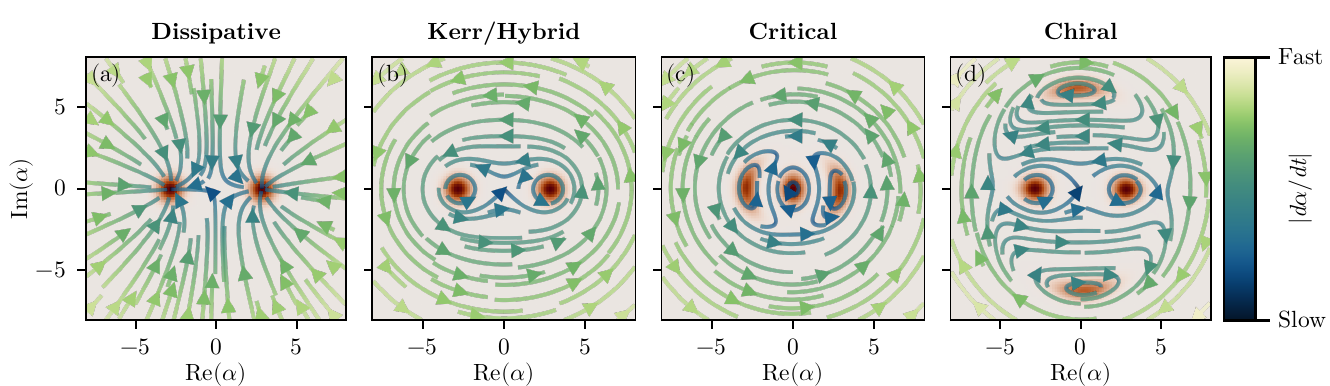}
    \caption{Phase-space representation of the manifold for (a) dissipative, (b) Kerr or hybrid, (c) critical, and (d) chiral cat states, and corresponding vector field obtained by the semiclassical analysis. While for a standard dissipative, Kerr, or hybrid cat errors can bring the manifold out of the bottom its potential, in the chiral case the system is attracted to the high-photon manifold. Parameters ($\mathrm{MHz}$) as in \cref{fig:main-sketch} with $\kappa_\phi = 10^{-4}\kappa_1$. Detuning and two-photon driving are chosen as to keep the code space with eight photons: (a) $\epsilon_s / 2\pi = 20$, (b) $\epsilon_s / 2\pi = 20$, (c) $\Delta/2\pi = -32$ and $\epsilon_s / 2\pi = 3.6$, and (d) $\Delta/2\pi = -4$ and $\epsilon_s / 2\pi = 13.2$.}
    \label{fig:semiclassical}
\end{figure*}

Detuning can trigger bistability.
Indeed, it is possible to find regions of the parameter space where the minimal bit-flip rate occurs when the system is bistable, with a vacuum-like steady state [see \cref{fig:main-sketch}(b)].
In driven-dissipative nonlinear systems, bistability occurs when the system admits multiple long-lived (almost-stationary) states, each characterized by a different photon number.
Its emergence is linked to a balance between the various parameters of the system: on the one hand, detuning and nonlinearity determine the energy of a photon, while drive and dissipation compete to populate or empty it.
To understand this phenomenon, one can again resort to a semiclassical  approximation that in this case reads
\begin{equation}
    \begin{split}
        \dot{\alpha}(t) =&  - \left(\frac{\kappa_1}{2} - \kappa_2 \abs{\alpha (t)}^2\right)\alpha (t) \\
        &- i 2 \epsilon_s \alpha^* (t) - i \alpha (t) \left(-\Delta + K_2  \abs{\alpha(t)}^2 \right)
    \end{split}
\end{equation}
This nonlinear equation admits, thanks to the addition of detuning, multiple solutions for the steady-state equation $\dot{\alpha} =0$ for an appropriate choice of parameters.
A stability analysis then finds that both the vacuum $\alpha =0$ and a high-photon manifold $|\alpha| \neq 0$ can coexist [see \cref{fig:main-sketch}(b)].
Once quantum fluctuations are re-introduced, these solutions become metastable. 
The steady state according to full quantum solution will display an abrupt change between the two states around a critical point, connected to the presence of a dissipative phase transition \cite{minganti2018,Berdou2023,Beaulieu2025,minganti2023dissipativephase}.
Around the critical point, the system will switch between the vacuum and the populated phase at random times. Instead, on either side of the transition, the system will remain for long time in one of the metastable manifolds to then irreversibly decay to the steady state \cite{macieszczak2016towards,macieszczak2021theory}.
There is a profound connection between second-order phase transitions and quantum information encoding \cite{lieu2020,Liu2024,lieu2025viewingfluxoniumlenscat}, that we further extend here to first-order ones.

A cat operated in this latter regime, called a \textit{critical cat} \cite{gravina2023criticalcat}, can thus still preserve quantum information on long timescales, even if the system eventually reaches the vacuum. 
Operating this critical cat in the metastable regime requires some specific procedures \cite{gravina2023criticalcat}.
When comparing the performance of the critical cat with an hybrid cat at $\Delta =0$, we see an improvement not only in the presence of one-photon dissipation, but also in the presence of dephasing as shown in \cref{fig:Kerr_vs_dissipative}.
As the hybrid cat was already outperforming the Kerr cat, we conclude that the critical cat selects the optimal parameters to encode quantum information in a Kerr dominated cat.
Finally, we observe that the system is again capable of exponentially suppressing errors, making it viable even in the presence of large dephasing.


When investigating the phase-space structure of the critical cat, as done in Fig.~\ref{fig:semiclassical}(c) we observe how it adds a layer of complexity compared to dissipative and Kerr cats, with detuning deforming the shape of the orbits, thus requiring larger fluctuations for the system to show bit flips.
In the limit of large detuning where the critical cat is operated, however, the vacuum becomes the attractor of the dynamics, implying that, eventually, a system will decay to it.

\subsection{The chiral cat}

Let us now include $K_3 \neq 0$.
In the idealized cat configuration, where $\kappa_1 = \kappa_{\phi} = \Delta =0$
also this term prevents the steady state from defining a decoherence-free subspace.
However, in the presence of photon loss and dephasing the inclusion of $K_3$ can improve the cat's performance, in analogy to what discussed for detuning in the critical cat.
%
%

\begin{figure}[b]
    \centering
    \includegraphics[width=\linewidth]{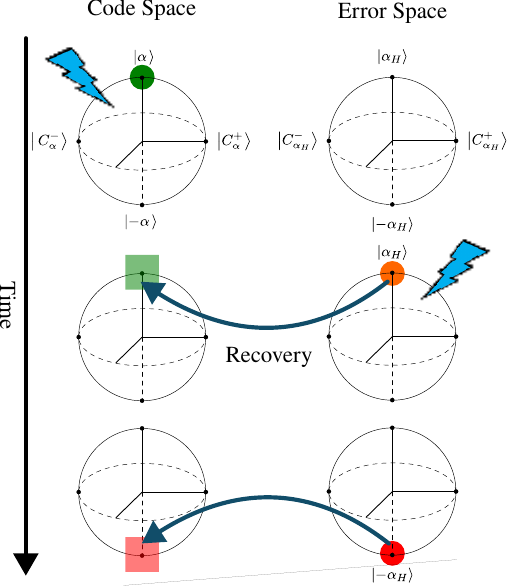}
    \caption{Within a coarse-grained quantum trajectory picture, in order for the system to display a logical error, two consecutive processes must occur. First, the state has to pass from the code space to the error space. Such an event, although improbable, occurs on a timescale that is not the shortest one in the problem. Then, a second jump in the error manifold must take place. This second jump is significantly more improbable. Error correction can then be performed by detecting if the qubit jumped to the error manifold, and in that case by reverting to the code space.}
    \label{fig:error-path}
\end{figure}

As the competition between $\Delta$ and $K_2$ generates bistability, so does that between $K_2$ and $K_3$.
This can be straightforwardly argued from the equation of motion of the mean field, which now reads~\cite{minganti2023dissipativephase}
\begin{equation}\label{eq:mean_field_chiral}
\begin{split}
    \dot{\alpha}(t) =&  - \left(\frac{\kappa_1}{2} - \kappa_2 \abs{\alpha}^2\right)\alpha - i 2 \epsilon_s \alpha^* \\ &- i \alpha \left(-\Delta + K_2  \abs{\alpha}^2 + K_3  \abs{\alpha}^4  \right).
\end{split}
\end{equation}
We show the solution to this equation in \cref{fig:main-sketch}(c).
Here, if the relative sign of $K_2$ and $K_3$ is different, two metastable manifolds emerge: one spanned by low-photon states $\rho_{\pm\alpha} = \ketbra{\pm \alpha}$, and one by high-photon number ones $\rho_{\pm \alpha_H} =\ketbra{\pm \alpha_H}$.
Compared to the bistability between $\Delta$ and $K_2$, where the two metastable states are the cat manifold and the vacuum, here both manifolds can encode quantum information, and define cat states and qubits.

Suppose now that we take a state initialized in $\ket{+ \alpha}$.
In a generic bistable configuration this state can jump to $\ket{- \alpha}$, generating bit flip errors, but also explore the states $\ket{\pm \alpha_H}$.
However, by choosing $\Delta <0$, $K_2 <0$, and $K_3>0$,
it is possible to suppress some of these transitions.
The mechanism underlying this behavior can be understood again using the semiclassical analysis shown in \cref{fig:semiclassical}(d).
Any perturbation that takes the cat qubit out of the $\ket{+ \alpha}$ manifold will either bring it back to itself, or will mainly lead to a decay into $\ket{+ \alpha_H}$. The opposite passage from $\ket{+ \alpha} \to \ket{- \alpha_H}$ is far less probable, as well as $\ket{+ \alpha} \to \ket{- \alpha}$.
Similar relations hold for the state $\ket{- \alpha}$.
We conclude that the manifold spanned by $\ket{\pm \alpha_H}$ acts as a trap for bit flip events, and one can engineer a right-bottom/left-top decay structure in phase space, where the transitions are mainly $\ket{+ \alpha} \to \ket{+ \alpha_H}$ and $\ket{- \alpha} \to \ket{- \alpha_H}$, thus inducing a \textit{chiral} structure.

This emergent chirality allows error correction to be applied.
Indeed, calling $\ket{\pm \alpha}$ the code space and $\ket{\pm \alpha_H}$ the error space (as in \cref{fig:bloch_chiral}), one can determine whether the system is in the code or error space without acquiring any information on the phase of the field, as these states have different photon number.
This makes it possible to detect errors and correct them, sending back $\ket{\pm \alpha_H} \to \ket{\pm \alpha}$.

Coarse-graining the dynamics of the quantum system so that dephasing and photon loss errors accumulate to determine an overall bit- or phase-flip errors, we can visualize the process leading to an error in the logical space, as done in Fig.~\ref{fig:error-path}.
For a logical error to occur, one needs \textit{both} $\ket{\pm \alpha} \to \ket{\pm \alpha_H}$ \textit{and} $\ket{\pm \alpha_H} \to \ket{\mp \alpha_H}$.
We can say that this error correction protocol allows the error correction rate of the large cat $\ket{\pm \alpha_H}$, while maintaining the phase flip error rate of the small encoding $\ket{\pm \alpha}$. Further error correction capability comes from the squeezed nature of the states in $\ket{\pm \alpha_H}$. This code thus permits an efficient correction of bit-flip errors in Kerr cat even in the presence of large dephasing.
The rest of the paper is dedicated to rigorously demonstrating this statement.



\section{Performance of the chiral cat}\label{sec:implementation_chiral_cat}

To determine the optimal regime of operation for the chiral cat qubit, we first assume that the system admits coherent states as attractor of the dynamics.
Namely, the system is either in $\ket{\pm  \alpha}$ or $\ket{\pm  \alpha_H}$. We then imagine a protocol where, if the system is detected in $\ket{\pm  \alpha_H}$, we project it back onto the corresponding $\ket{\pm  \alpha}$; otherwise no operation is applied.
Overall, the measurement and recovery procedure takes the form of the projector $\hat{\Pi}_{\alpha_H} = \ketbra{\alpha_H} + \ketbra{-\alpha_H}$ and recovery 
$\hat{R} = \ketbra{\alpha}{\alpha_H} + \ketbra{-\alpha}{-\alpha_H}$, that combined give
\begin{equation}\label{eq:recovery_operator}
\begin{split}
    &\hat{R} \, \hat{\Pi}_{\alpha_H} + (\hat{\mathbb{1}} - \hat{\Pi}_{\alpha_H})  = 
    \\
    & \quad  \ket{\alpha}\left(\bra{\alpha} + \bra{\alpha_H} \right) +  \ket{-\alpha}\left(\bra{-\alpha} + \bra{-\alpha_H}\right).
\end{split}
\end{equation}

\begin{figure}[t]
    \centering
    \includegraphics[width=\linewidth]{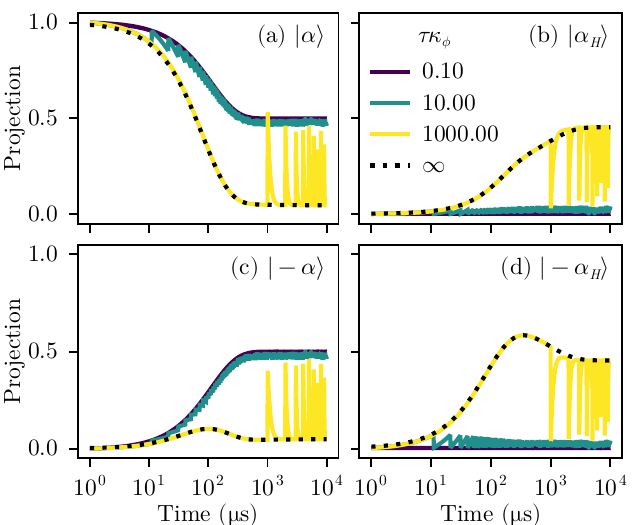}
    \caption{Example evolution of a state $\ket{\alpha}$ with the recovery operation applied at various recovery times $\tau$. Without correction (dotted black with $\tau = \infty$) the population on the left lobe is driven towards the high-manifold and ends in an equal classical mixture of $\ket{\pm \alpha_H}$. Similarly, if correction is too slow, the recovery fails at preserving the code space. Instead, for $\tau\kappa_\phi \ll 1000$, the code space is preserved and the decay towards the steady state $\ketbra{\alpha} + \ketbra{-\alpha}$ is slowed down.
    Parameters: $\kappa_1/2\pi = \SI{5}{kHz}$, $\kappa_\phi = 10 \kappa_1$, $\epsilon_s/2\pi = \SI{59}{MHz}$, $\Delta/2\pi = \SI{82}{MHz}$, $K_2/2\pi = -\SI{42}{MHz}$, $K_3/2\pi = \SI{2}{MHz}$ and $\kappa_2 = 0.01\abs{K_2}$.}
    \label{fig:recovery}
\end{figure}

While this intuition is valid for the coherent-state approximation and if we assume at all times to be either in $\ketbra{\pm \alpha}$ or $\ketbra{\pm \alpha_H}$, the mixed nature of the encoding and the fact that the system will also explore other states require being more mathematically careful.
To generalize this idea to the actual regime of operation, we call $\rho_{\pm\alpha}$ the coherent-like states associated with the low-photon number manifold,
and $\rho_{\pm\alpha_H}$ that of the high-photon number states.
Periodically, we detect whether the system transitioned to the high-photon manifold. To do that, we select the most probable states of $\rho_{\pm\alpha_H}$ and construct the projector onto them, which we call $\hat{\Pi}_{\alpha_H}$ \footnote{Operationally, we diagonalize $\rho_{\pm\alpha_H} = \sum_j p_j^{(\pm)} \ketbra*{\psi_j^{(\pm)}}$ and define $\hat{\Pi}_{\alpha_H} = \sum_{j: \, p_j>10^{-4}} \sum_k \ketbra*{\psi_j^{(k)}}$.
We verified that changing the threshold to smaller values $p_j$ does not qualitatively change the results.}.
If the system is not in the $\rho_{\pm\alpha_H}$ manifold, we do not perform any error correction.
Otherwise, we act with an idealized recovery operation $\hat{R}$ sending each of the components of the state $\rho_{\pm\alpha_H}$ to the most probable one in $\rho_{\pm\alpha}$.
Namely, the overall recovery procedure reads
\begin{equation}\label{eq:recovery_superoperator}
    \rho_R = \mathcal{R} \rho = \hat{R} \, \hat{\Pi}_{\alpha_H} \, \rho \,\hat{\Pi}_{\alpha_H}^\dagger  \, \hat{R}^\dagger  + (\hat{\mathbb{1}} - \hat{\Pi}_{\alpha_H}) \rho  (\hat{\mathbb{1}} - \hat{\Pi}_{\alpha_H}^\dagger).
\end{equation}

\begin{figure*}[t]
    \centering
    \includegraphics[width=\linewidth]{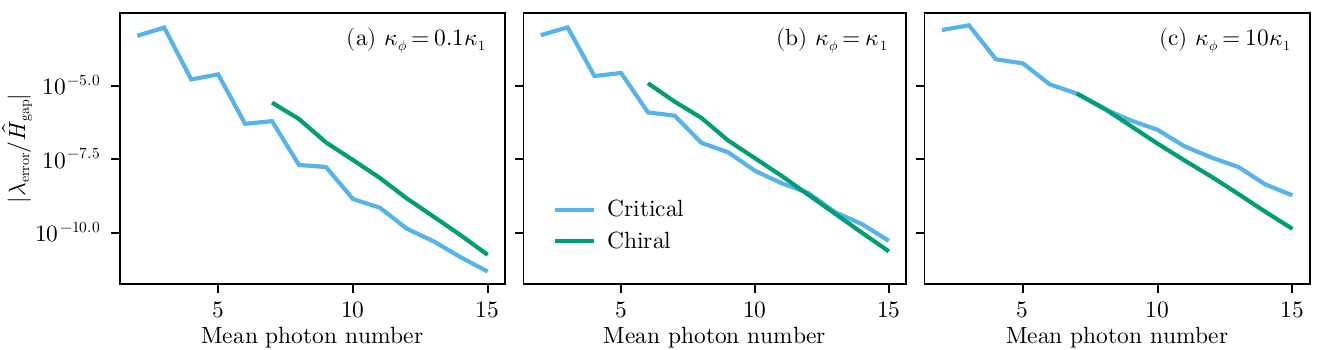}
    \caption{Results of the parameter optimization for both critical (blue) and chiral (green) cats for increasing values of dephasing: (a) $\kappa_\phi /\kappa_1 =0.1$, (b) $\kappa_\phi /\kappa_1 =1$, and (c) $\kappa_\phi /\kappa_1 =10$. As a function of the mean photon number of the code space, we show the ratio between the error rate (the effective error rate obtained by applying detection and recovery protocol to the chiral cat) with respect to the Hamiltonian gap. For the chiral cat we find an exponential scaling of the form $\sim \exp(-\gamma n)$ where $\gamma$ is $0.71$, $0.54$ and $0.48$ for each value of $\kappa_\phi$ respectively. This is attained at an almost constant Kerr nonlinearity of $45\cdot 2\pi\mathrm{MHz}$ and $K_3$ of the order $\SI{}{2\pi MHz}$. Optimization $\Delta$, $K_2$ and $\epsilon_s$ with fixed $\abs{\kappa_2/K_2} = 0.01$ and $\kappa_1/2\pi = \SI{5}{kHz}$.}
    \label{fig:scale_ratio_photon_number}
\end{figure*}

To find the optimal working point for this algorithm, we then search for the point where the system admits the chiral structure with four modes. 
Then we check that the passage $\ket{\alpha} \to \ket{\alpha_H}$ is slow, but still faster than $\ket{\alpha} \to \ket{-\alpha}$.
This translates into investigating three Liouvillian eigenvalues: (i) $\lambda_{\rm leak}$ that describes the leakage rate of the whole code space, spanned by the metastable manifold $\rho_{\pm \alpha}$, into the error space spanned by $\rho_{\pm \alpha_H}$;
(ii) $\lambda_{\rm chiral}$ that describes how
$ \ket{\pm\alpha} \to \ket{\pm \alpha_H}$ occurs;
(iii) $\lambda_{H}$ that describes the passage $\ket{\pm\alpha_H} \to \ket{\mp \alpha_H}$. 
While this restriction may provide a suboptimal solution to the problem, all these elements can be automatically constructed using the eigendecomposition of the Liouvillian, and the automated search discussed in the \cref{apx:optimization}.

We then assume to monitor the system at a certain rate $\tau$ and we perform the recovery operation $\mathcal{R}$ at this frequency, as shown in \cref{fig:recovery}.
Therefore, the time evolution of the system is of the form
\begin{equation}
    \rho(t+ \tau) = \mathcal{R} e^{\lv \tau} \rho(t).
\end{equation}
The optimal performance of the code is then found by investigating the bit-flip error rate sending $\tau \to 0$.
In this limit, we can define an effective error rate $\lambda_{\rm error}$ describing the rate at which the bit flip $\ket{\alpha} \to \ket{- \alpha}$ occurs.
In practice, we verified that taking $\kappa_\phi \tau \ll 1 $ does not significantly change this rate.

We show the performance of this chiral cat in suppressing bit flip errors and compare it to a critical (i.e., an optimal hybrid detuned) cat in \cref{fig:scale_ratio_photon_number}.
To provide a fair comparison, we compare the bit flip rate to the Hamiltonian gap between the ground and first excited state $E_1 - E_0$, with $E_j$ the Hamiltonian energies. 
This quantity captures the speed at which Hamiltonian operations can be performed.
While in photon-loss dominated configurations the chiral cat matches the performance of the hybrid one (demonstrating that higher-order nonlinearies are not necessarily detrimental), for dephasing dominated systems, the chiral cat outperforms the critical cat.

\subsection*{Gates}

Since the chiral cat behaves as a Kerr cat in the code space, gates can be directly adapted from those of standard Kerr cats \cite{puri2020bias}.

\begin{figure}[t]
    \centering
    \includegraphics[width=\linewidth]{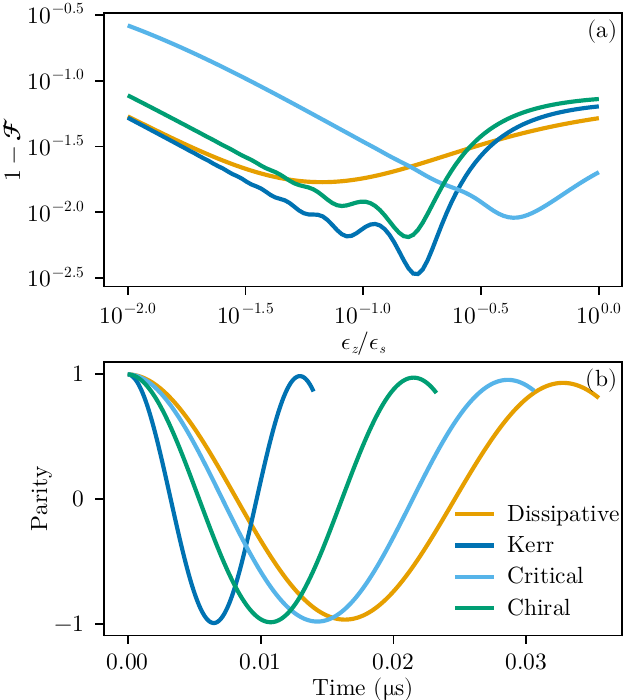}
    \caption{Performance of $z$ gate in a dissipative (orange), Kerr (dark blue),  critical (light blue), and chiral (green) configuration. (a) As a function of the ratio $\epsilon_z/\epsilon_s$, the error in performing the $z$ gate. (b) As a function of time, parity oscillations at the optimal $\epsilon_z$ point. Parameters as in \cref{fig:semiclassical}.}
    \label{fig:zgate}
\end{figure}

A gate that may be adversely affected by the presence of $K_3$ is the $z$ gate, which has form
\begin{equation}
    \hat{H}_z = \epsilon_z (\hat{a} + \hat{a}^\dagger).
\end{equation}
Indeed, the presence of the drive may induce transitions between the low-and high manifold.
This is not the case, as we show in Fig.~\ref{fig:zgate}. 
We demonstrate the gate can be performed with an efficiency similar to that of the Kerr cat limit [Fig.~\ref{fig:zgate}(a)].
Compared to a dissipative cat, we see that the operation can be performed much more rapidly, and with a significantly smaller error rate.
We conclude that the chiral cat maintains the capability of Kerr cat to be rapidly operated [Fig.~\ref{fig:zgate}(b)].

\section{Implementation and concatenation}

\begin{figure}[b]
    \centering
    \includegraphics[width=\linewidth]{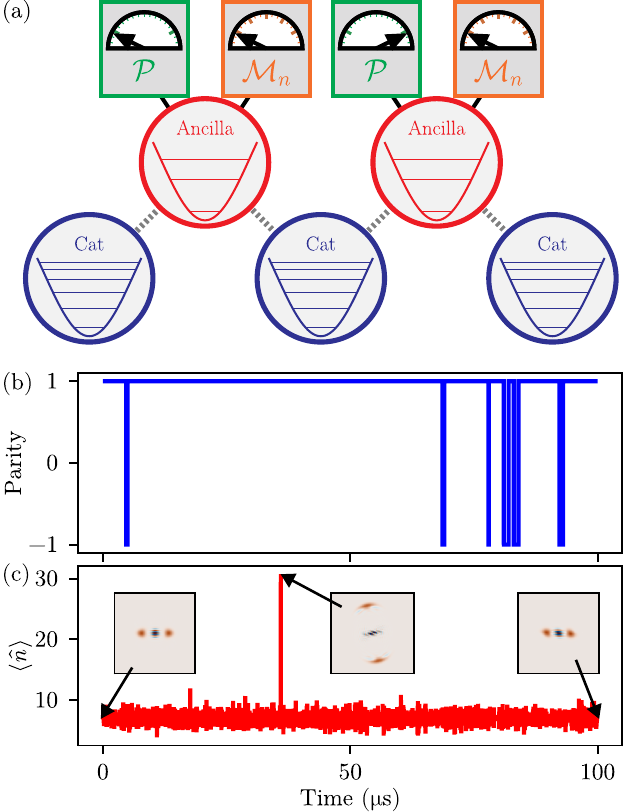}
    \caption{(a) Repetition code with cat data and ancillas qubit. The latter are used both to perform parity measurement (green boxes) and photon number discrimination measurement (orange boxes). Upon the detection of an error either in parity or in photon number, the appropriate error correction protocol is applied. Along a single quantum trajectory and as a function of time evolution of (b) parity and (c) photon number. A jump in parity corresponds to a phase flip error in one qubit, which can be corrected by the parity check performed by the repetition code. Parity jumps do not affect the photon number. 
    A jump in the photon number, instead, corresponds to the $\ket{\pm\alpha} \to \ket{\pm\alpha_H}$ transition.
    As it does not yet correspond to a bit flip, this error can be detected and corrected, as shown in the trajectory.
    Notice that this jump does not directly entail a jump in the parity. Parameters as in \cref{fig:recovery}.
    }
    \label{fig:concatenated_code}
\end{figure}

Having demonstrated the capability of the code in correcting for bit-flip events in an idealized scenario, here we discuss how to actually perform their correction, while simultaneously correcting for phase-flip ones.
The basic strategy we propose, shown in Fig.~\ref{fig:concatenated_code} is to perform a repetition code using nonlinear ancillas, as in the experimental realization of a repetition code done in Ref.~\cite{Putterman2025}.
The role of the ancilla can be dual: they can be used to perform the parity checks needed to detect phase flip errors and to detect the $\ket{\alpha} \to \ket{\alpha_H}$ chiral errors.
For example, the latter could be checked using a dispersive readout protocol~\cite{blais2021cqe}.
We also stress that in configurations where two-photon dissipation is achieved through an additional buffer mode, the emission spectroscopy of the buffer also carries information on the state of the memory.
 
\subsection{Simultaneous detection of bit and phase flip errors}

We consider here a simplified architecture, where we assume the ancilla to perform a parity and photon-number readout without any errors.
We then simulate a repetition code with 7 logical qubits.
Upon detection of a bit flip, we apply the idealized recovery protocol described above.
Similarly, if a phase flip is detected, we correct for the parity error by injecting a photon.

In Fig.~\ref{fig:concatenated_code}, we show a quantum trajectory for a system initialized in an even cat, where several phase flip events occur, as well as the jump from $\ket{\alpha} \to \ket{\alpha_H}$.
As a jump in parity [Fig.~\ref{fig:concatenated_code}(a)] does not change the photon number, a jump in photon number does not affect parity [Fig.~\ref{fig:concatenated_code}(b)].
The Wigner functions, shown at various time along the dynamics, clearly demonstrate the capability of the code to store and preserve quantum information over long times.

It should be noticed that the chiral error rate $\ket{\alpha} \to \ket{\alpha_H}$ also affects the performance of the repetition code.
Indeed, the phase flip error rate is significantly larger in the $\alpha_H$ manifold and a chiral jump will effectively reduce the distance of the error correction code, removing one of the qubits from the repetition array.
In practice, calling $p_{\rm chiral} $ the probability of a qubit to jump in the time interval $\tau$ of one detection and recovery cycle, we can model the overall phase-flip error rate as
$p_{\rm phase-flip} \simeq \kappa_1 |\alpha|^2 \tau  (1 + p_{\rm chiral}  |\alpha_H|^2/|\alpha|^2)$.





\subsection{Recovery protocol}

\begin{figure*}
    \centering
    \includegraphics[width=\linewidth]{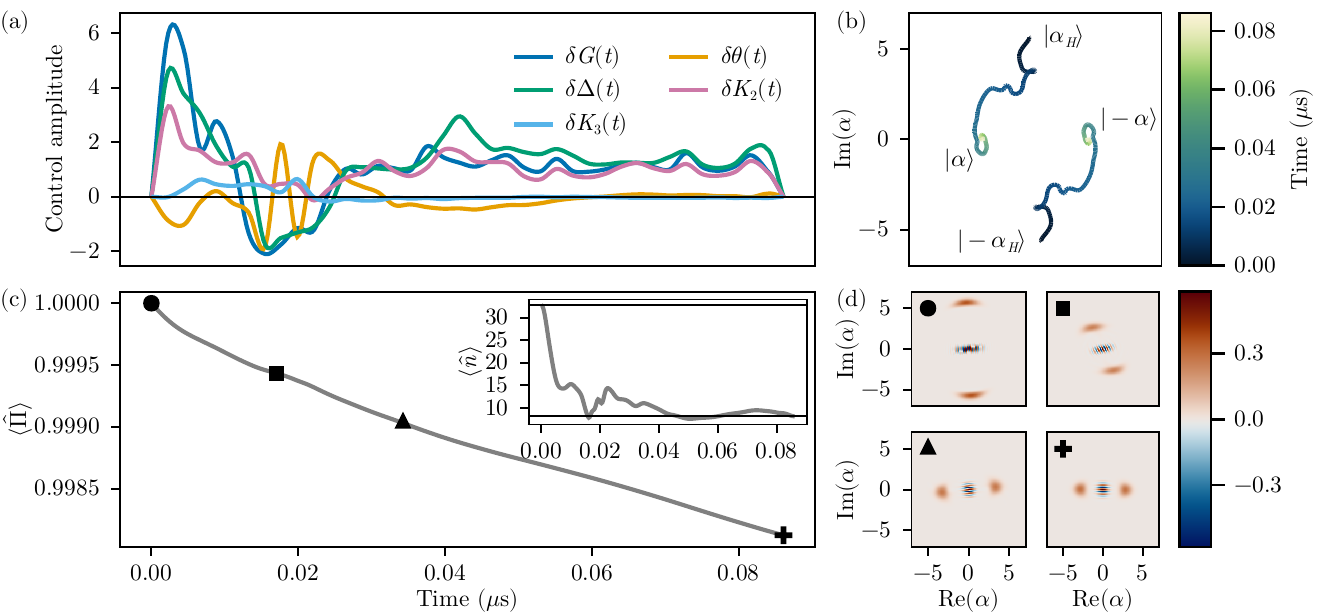}
    \caption{Optimal-control-based adiabatic recovery protocol. (a) Time evolution of the optimized system parameters, expressed as relative variations with respect to their idle values. The parameters are tuned smoothly in time via gradient-descent optimization to maximize the fidelity of the mapping $\ket{\pm\alpha_H} \rightarrow \ket{\pm\alpha}$. (b) Time evolution of $\alpha(t) = \expval{\hat{a}}(t)$ in a phase-space representation for a state initialized in $\ket{\pm \alpha_H}$. (c) Evolution of the state parity $\langle \hat{P} \rangle$ for the system initialized in $\ket{\mathcal{C}^{+}_{\alpha_H}}$. The parity remains nearly constant throughout the protocol, indicating that the recovery preserves the coherence between the components of the cat state. In the inset, we show the evolution of the photon number. (d) Wigner functions along the evolution for a system initialized in an even cat state at high-photon number $\ket{\mathcal{C}^{+}_{\alpha_H}}$. For this choice of parameters the procedure achieves a final fidelity of $\mathcal{F} = 99.3\%$, demonstrating high-fidelity correction with minimal degradation of the encoded qubit’s phase coherence. Parameters as in \cref{fig:recovery}.}
    \label{fig:recovery-optimal-control}
\end{figure*}

Up to this point, we considered an idealized version of the detection and recovery protocol. 
As we briefly discussed above, we can use a dispersively coupled transmon to perform a photon-distribution detection and determine if a jump from the code to the error manifold has occurred.
Given the significant change in photon number (from $n \approx 10$ to $n\approx 30$ in the example considered in this manuscript), and the state of control of transmon-based devices \cite{blais2021cqe,Putterman2025,google2024}, we argue that such a detection does not provide a major limitation to the correction protocol.

As for the recovery procedure, being at the edge of optically bistable processes, we notice that relatively small changes in the parameters can be engineered to introduce large changes in the system's state \cite{minganti2023dissipativephase}.
We show here that such a property can be leveraged to perform the desired chiral correction procedure, using an approach based on optimal control of the system's nonlinearities and drives.
Such a task can be accomplished, for instance, when operating a flux-tunable SNAIL-like device \cite{FrattiniPRApp2018,grimm2020,frattini2022,Frattini2024} via time-tunable external fluxes.

We perform an optimization to find the optimal recovery protocol. The goal of the optimization is to map $\ket{\pm \alpha_H} \to \ket{\pm \alpha}$ with maximal fidelity, while only marginally degrading the parity of the state along the recovery.
We define a time-dependent Liouvillian $\lv(t)$ and discretize the time $t \in [0, t_{\rm end}]$ to $N$ time steps. At each time step, the parameters $\epsilon_d(t)e^{i\theta(t)}$, $\Delta(t)$, $K_2(t)$, and $K_3(t)$ are optimized, and the final pulse shapes are the interpolation of these optimized points. 
We employ a gradient descent method to optimize the time dependence of the system's parameters for a system initialized in $\ket{\alpha_H}$.
As a cost function we take the fidelity between $\rho_\alpha$ and $\rho(t_{\rm end})$, where $\rho(t_{\rm end})$ is the state at the end of the protocol. 
We fix $t_{\rm end}$ to $\SI{0.084}{\mu s}$. As we detail below, such a protocol guarantees that parity is only marginally degraded by the recovery, as the photon number rapidly drops to the final required value, and the protocol time is short enough to prevent one-photon dissipation from degrading parity. 
The symmetry of the system guarantees that the same protocol will map $\ket{-\alpha_H} \to \ket{-\alpha}$.

In \cref{fig:recovery-optimal-control} we show an example of such an optimal-control-based protocol. First, in \cref{fig:recovery-optimal-control}(a) we plot, as a function of time, the optimized relative change of the system's Hamiltonian parameters with respect to the ones used when the cat is idle. 
We highlight the relatively small changes as a function of time.
Performing this recovery procedure, we find that a state initialized in $\ket{\pm \alpha_H}$ reaches the corresponding state $\ket{\pm \alpha}$ with a fidelity of $99.3\%$.
We visualize the change in the state as a function of time by plotting $\expval{\hat{a}}(t)$ in  \cref{fig:recovery-optimal-control}(b).

In \cref{fig:recovery-optimal-control}(c) we then plot the parity as a function of time for a system initialized in $\ket{\mathcal{C}^{+}_{\alpha_H}}$. The recovery procedure only marginally degrades the parity of the system, with $\langle\hat{P}\rangle > 0.998$. In the inset of \cref{fig:recovery-optimal-control}(c) we see that, indeed, the photon number rapidly drops to the idle value, ensuring only a minor degradation of parity.
Finally, the  Wigner function of an even cat $\ket{\mathcal{C}^{+}_{\alpha_H}}$ is faithfully restored to that $\ket{\mathcal{C}^{+}_{\alpha}}$, as shown in \cref{fig:recovery-optimal-control}(d).
We conclude that this optimal control protocol would allow performing the wanted error correction of bit-flips without further degrading phase flips. 

Although we have limited ourselves to dynamically changing the system's non-linearities and drive parameters, more advanced error correction procedures can be envisioned.
For instance, engineered photon-number-dependent high-order dissipation could be tailored to perform active or autonomous error correction, preventing the system from reaching the error manifold and confining it to the low-photon code space.


\section{Conclusions}
We introduced a chiral cat qubit, a logical qubit that works in the presence of higher-order nonlinearities.
We show how these higher-order nonlinear processes can be harnessed to improve the performance of the logical qubit, providing a pathway to protect Kerr cats from dephasing. 
The system we consider displays optical bistability, i.e., the simultaneous presence of multiple solutions according to the semiclassical approximation. 
We show that these states can form a logical code and error space, 
and we demonstrate how the chiral structure acquired by the phase space can be used for error correction. 
Our work shows that the chiral cat has the desired error correction feature, while also maintaining the capability of Kerr cats to be rapidly operated. At the same time, the chiral cat is characterized by exponential error correction capabilities in the size of the cat, even in the presence of large dephasing.
We provide a pathway based on straightforward optimal control to achieve the wanted performance.

Beyond the specific example investigated here, our work demonstrates how the rich landscape of parametric, critical, and topological phenomena in open bosonic systems can endow bosonic codes with error correction capabilities.
We plan to investigate how chiral properties can be applied to different driving schemes, or investigate colored dissipation and autonomous stabilization possibilities for chiral codes.

\begin{acknowledgments}
    We thank Filippo Ferrari, Luca Gravina, and Roberta Zambrini for insightful discussions.
    FM acknowledges insightful discussion with team members at Alice \& Bob, in particular Giulio Campanaro, Joachim Cohen, Andr\'e Melo, and S\'ebastien Jezouin.
    We acknowledge the Spanish State Research Agency, through the Mar\'ia de Maeztu project CEX2021-001164-M, through the COQUSY project PID2022-140506NB-C21 and -C22, and through the QuantCom project CNS2024-154720, all funded by MCIU/AEI/10.13039/501100011033; the project is funded under the Quantera II programme that has received funding from the EU’s H2020 research and innovation programme under the GA No 101017733, and from the Spanish State Research Agency, PCI2024-153410 funded by MCIU/ AEI/10.13039/50110001103;  MINECO through the QUANTUM SPAIN project, and EU through the RTRP - NextGenerationEU within the framework of the Digital Spain 2025 Agenda. V.S. acknowledges support by the Swiss National Science Foundation through Projects No. 200020\_185015, 200020\_215172, 200021-227992, and 20QU-1\_215928.
\end{acknowledgments}

\appendix

\begin{figure*}
    \centering
    \includegraphics[width=\linewidth]{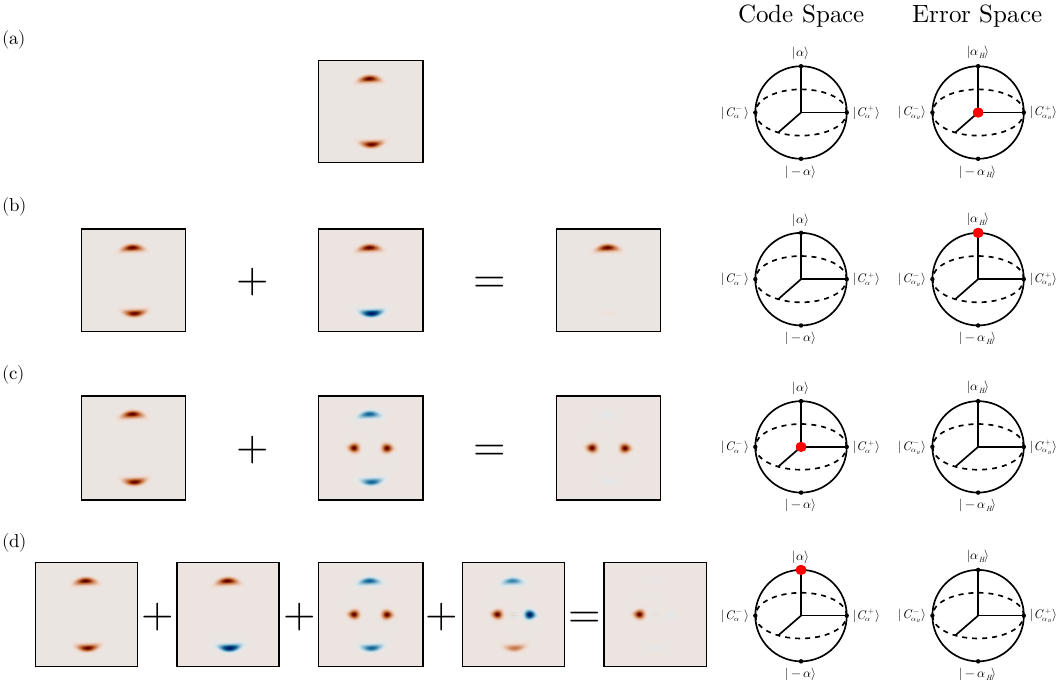}
    \caption{Construction of the logical states using the Liouvillian eigenvalues.
    On the left we plot the Wigner function of the Liouvillian eigenoperators. In the center the resulting state. On the right, their representation in the error and code space.
    (a) The steady state $\rho_{\rm ss}$.
    (b) The state associated with $\rho_{\alpha_H} \to \rho_{-\alpha_H}$ is $\rho_{\rm H}$. Added to $\rho_{\rm ss}$ gives $\rho_{\alpha_H}$.
    (c) The leakage matrix $\rho_{\rm leak}$ describes how the center of the Bloch sphere in the code space passes to the centre of the Bloch sphere in the error space.
    (d) Adding the chiral decay into the mixture, we finally obtain $\rho_\alpha$.
    }
    \label{fig:Liouvillian_eigendecomposition}
\end{figure*}

\section{Interpretation of the Liouvillian eigenoperators}\label{apx:lv_eigenoperators}

As we discussed in the main text, three Liouvillian eigenvalues and eigenoperators come into play when discussing the chiral cat error: (i) $\lambda_{\rm leak}$ and the corresponding $\rho_{\rm leak}$; (ii) $\lambda_{\rm chiral}$ and $\rho_{\rm chiral}$; and (iii) $\lambda_{H}$ and $\rho_{\rm H}$.
To understand why this is the case, and their role in determining the qubit performance, in \cref{fig:Liouvillian_eigendecomposition} we plot the various state that lead to the construction of the logical state $\rho_{+\alpha}$.
First, in \cref{fig:Liouvillian_eigendecomposition}(a) we plot the steady state of the system. 
This is of the form $\rho_{\rm ss} \simeq (\rho_{+\alpha_H} + \rho_{-\alpha_H})/2$.
To obtain the state $\rho_{+\alpha_H}$ we need to add the eigenstate $\rho_H$, as shown in \cref{fig:Liouvillian_eigendecomposition}(b).
Similarly, to pass from the error manifold to the centre of the Bloch sphere in the code manifold, we have $\rho_{\alpha} +\rho_{-\alpha} = \rho_{\rm ss} + \rho_{\rm leak}$, as shown in \cref{fig:Liouvillian_eigendecomposition}(c).
Finally, the state $\rho_{\alpha}$ is obtained by summing all these processes, so that, as shown in \cref{fig:Liouvillian_eigendecomposition}(d),
\begin{equation}
    \rho_\alpha = \rho_{\rm ss} + \rho_{\rm leak}  + \rho_H + \rho_{\rm chiral}.
\end{equation}
The time evolution of this state is of the form
\begin{equation}
    e^{\lv t}\rho_\alpha = \rho_{\rm ss} + e^{\lambda_{\rm leak} t}\rho_{\rm leak}  + e^{\lambda_{\rm H} t}\rho_H + e^{\lambda_{\rm chiral} t} \rho_{\rm chiral}.
\end{equation}
Remarkably, here there is no direct bit-flip error between $\rho_{\alpha}$ and $\rho_{-\alpha}$, but the passage to the high-photon manifold is always required.
Finally, we remark that $|\lambda_{H}|< |\lambda_{\rm leak}|<|\lambda_{\rm chiral}|$.

\section{Parameter optimization}\label{apx:optimization}

To realize a chiral cat, it is crucial to identify the optimal parameters, as chirality manifests within a narrow parameter regime. Moreover, the Liouvillian gap is no longer the quantity that determines the bit flip error rate (as in \cref{fig:Kerr_vs_dissipative}), but the chiral structure and its associated eigenoperators have to be properly accounted between the four slowest eigenmodes (see \cref{apx:lv_eigenoperators}). 
To extensively search for the optimal parameters (see \cref{eq:full_liouvillian}), 
we employed evolutionary minimization algorithms to determine the parameter set that yields the optimal bit-flip time using the correction method described in \cref{sec:implementation_chiral_cat}.
Such a choice was mainly motivated by the non-differentiable nature of the minimization function.
Specifically, we choose utilized the \texttt{Evolutionary Centers Algorithm} from \texttt{Metaheuristics.jl} \cite{metaheuristics2022}. This algorithm leverages genetic operations such as selection, crossover, and mutation to guide an initial population of $N_{\rm pop}$ individuals toward the global minimum over $N_{\rm iter}$ iterations \cite{goldberg1989genetic}. The following sections detail the optimization process to obtain the bit-flip time between $\ket{+\alpha} \leftrightarrow \ket{-\alpha}$ as a function of $\abs{\alpha}^2$.

\subsection{Search space}

The problem is characterised by seven parameters. We assume three of them to be fixed: the one-photon dissipation ($\kappa_1$), two-photon dissipation ($\kappa_2$), and dephasing ($\kappa_\phi$) rates.
While various relative ratios of $\kappa_1$ and $\kappa_\phi$ are explored,  $\kappa_2$ is chosen to be $1\%$ of the Kerr interaction $K_2$.
The others, detuning ($\Delta$), two-photon drive ($\epsilon_s$), Kerr ($K_2$) and the first high-order correction ($K_3$), are instead free parameters.  However, as we want to see the scaling with the mean photon number of the code space ($\abs{\alpha}^2$), we can use \cref{eq:mean_field_chiral} to fix the driving strength $\epsilon_s$ \cite{minganti2023dissipativephase}.
Note that the driving strength $\epsilon_s$ is not a free parameter:
as we fix the mean photon number of the code space ($\abs{\alpha}^2$) we can use the mean field equation \eqref{eq:mean_field_chiral} to it \cite{minganti2023dissipativephase}. 

All parameters are bounded by experimentally realizable values. While the Kerr nonlinearity can be tuned to tens of MHz \cite{puri2017,grimm2020}, for the higher-order correction we have no clear literature comparison.
As we do not want to fix a particular superconducting element to operate the chiral cat, we use the transmon model to bound $K_3$, and possible higher-order order corrections. 
The transmon Hamiltonian is given by $\hmt = 4 E_C \opn^2 - E_J \cos\hat{\phi}$, where $\hat{\phi} = \inpar{\frac{2 E_C}{E_J}}^{1/4}(\opad + \opa)$ is the phase operator, $E_C$ the charge energy and $E_J$ the Josephson energy \cite{koch2007transmon,blais2021cqe}. Expanding the cosine term, and discarding with different number of creation and annihilation operators, leads to the Hamiltonian $\hmt = \sum_m K_m \opad[m] \opa^m$ where the expressions can be found by normal ordering the annihilation and creation operators, setting $K_1 = \Delta$.
For instance, we have \footnote{The other terms up to order five are:
\begin{align}
   K_4 \approx& E_J \left[ - \frac{70}{8!}\inpar{\frac{2E_C}{E_J}}^{2} + \frac{3150}{10!}\inpar{\frac{2E_C}{E_J}}^{5/2} \right] \\
    K_5 \approx& E_J \left[ \frac{252}{10!}\inpar{\frac{2E_C}{E_J}}^{5/2} \right] \ .
\end{align}}
\begin{align}
    K_2 \approx& -\frac{E_C}{2} + E_J \left[ \frac{90}{6!}\inpar{\frac{2E_C}{E_J}}^{3/2} \right. \nonumber \\ 
    &- \left. \frac{1260}{8!}\inpar{\frac{2E_C}{E_J}}^{2} + \frac{18900}{10!}\inpar{\frac{2E_C}{E_J}}^{5/2} \right] \\
    K_3 \approx& E_J \left[ \frac{20}{6!}\inpar{\frac{2E_C}{E_J}}^{3/2} \right. \nonumber \\
    &- \left. \frac{560}{8!}\inpar{\frac{2E_C}{E_J}}^{2} + \frac{12600}{10!}\inpar{\frac{2E_C}{E_J}}^{5/2} \right] 
\end{align}
We note that the cosine potential already ensures an alternating sign between each term, needed to trigger multistability and  chirality. Higher-order corrections are also significantly smaller than the $K_2$ and $K_3$. 
Nonetheless, we include them in our optimization protocol to demonstrate the robustness of chirality to the presence of higher-order terms.
Notice that including these do not increase the search space, that still requires determining $E_C$ and $E_J$ rather then $K_2$ and $K_3$. 
Finally, the bounds for the optimization are 
\begin{center}
    \begin{tabular}{c|cc}
        \toprule
        Parameter & Min & Max \\ \midrule
        $\Delta$ (MHz/$2\pi$) & $-1000$ & $0$ \\
        $E_C$ (MHz/$2\pi$) & $1$ & $100$ \\
        $E_J/E_C$ & $5$ & $150$ \\ \bottomrule
    \end{tabular}
\end{center}
resulting in $K_2/2\pi \in [-50, -0.5]$ MHz and $K_3/2\pi \in [0.07, 4]$ MHz.

\subsection{Minimization function}

The target quantity to be minimized is the left-right transition rate once the correction operation introduced in \cref{sec:implementation_chiral_cat} is used. However, chirality only occurs in a very small parameter regime of the entire search space. Therefore, for each parameter set, we need to identify if there is a chiral mode (see \cref{apx:lv_eigenoperators}), extract the density matrices of both $\rho_{\pm \alpha}$ and $\rho_{\pm \alpha_H}$, construct the recovery operation [\cref{eq:recovery_operator}], and finally, determine the new transition rate. For this, we will use the fact that the eigenoperators describing the code and error spaces are metastable, resulting in a separation in the Liouvillian spectrum that distinguishes the four slowest eigenmodes from the rest. 
We also check that the vacuum in not the steady state, ensuring that all dynamics are contained within the manifold spanned by $\{\rho_{\pm \alpha}, \rho_{\pm \alpha_H} \}$. Below, we describe how this process is performed.

\paragraph{Detection of chiral eigenoperator} The chiral mode is an antisymmetric operator which, if it exists, lives in the second symmetry sector [$k=-1$ in \cref{Eq:Eigenspectrum}]. Moreover, since the chiral mode should be long-lived, we can restrict the search to only the two slowest eigenmodes. We know that this mode is decomposed as $\rho_{\rm chiral} = c_1 (\rho_{\alpha} - \rho_{-\alpha}) - c_2(\rho_{\alpha_H} - \rho_{-\alpha_H})$ where $c_1$ and $c_2$ are two proportionality constants resulting from numerical diagonalization. A direct method would be to eigendecompose the operator and filter the eigenvectors depending on the position in phase space. However, this leads to incorrect results, since the high-photon manifold may not lie on the imaginary axis in the Wigner function. Hence, we used a different approach which does not depend on the actual position of the lobes. We calculate the Wigner function $W_{\rm chiral}$ of $\rho_{\rm chiral}$ which results in a matrix that is mostly zero except in around $\rho_{\pm \alpha}$ and $\rho_{\pm \alpha_H}$. Then, we calculate the contour lines for a given value $\ell$ of the normalized matrix $\abs{W_{\rm chiral}}$. The chiral mode is identified if only four contour lines are found. We then check the relative sign the Wigner function within those contour [c.f. \cref{fig:Liouvillian_eigendecomposition}(d)]. 
During the optimization we fixed $\ell = 0.05$.

\paragraph{Extraction of density matrices} Given the four slowest eigenmodes, we use the procedure outlined in \cref{apx:lv_eigenoperators} to construct the four density matrices. 

\paragraph{Construction of the recovery operation} In general, the low manifold is well approximated by the pure coherent states $\ket{\pm \alpha}$ but this is not the case for the high manifold. In order to construct the projection and recovery operation we eigendecompose $\rho_{\pm \alpha_H}$ to obtain $\{(\lambda_j^{\pm \alpha_H}, \ket{v_j^{\pm \alpha_H}})\}_j$ where, due to the symmetry, $\lambda_j^{\alpha_H} = \lambda_j^{- \alpha_H}$ and the eigenvectors are equal up to a rotation. Only a few eigenvalues are relevant in the decomposition so we set a threshold value $\lambda_{\rm th} = 10^{-4}$. Hence, the projection and recovery operators are
\begin{eqs}
    \hat{\Pi}_{\alpha_H} &=  \sum_{\lambda_j^{\pm \alpha_H} > \lambda_{\rm th}} \op{v_j^{+ \alpha_H}} + \op{v_j^{- \alpha_H}} \\
    \hat{R} &=  \sum_{\lambda_j^{\pm \alpha_H} > \lambda_{\rm th}} \op{\alpha}{v_j^{+ \alpha_H}} + \op{-\alpha}{v_j^{- \alpha_H}}
\end{eqs}
This guarantees that any component of a state in the high manifold is taken to the low manifold.

\paragraph{Calculation of transition rate} To efficiently calculate the transition rate, we employ the theory of classical metastability \cite{macieszczak2021theory}, which allows us to describe the quantum jumps as classical transitions. This approach enables us to map the full Hilbert space into a four-dimensional manifold spanned by the four lobes. We begin by determining the left eigenoperators of the Liouvillian, defined as $\lv^\dagger \sigma_j = \lambda_j^* \sigma_j$, and sorting them in the same order as the right eigenoperators. We then calculate the coefficient matrix $[\hat{C}]_{j,\beta} = \tr \sigma_j^\dagger \rho_\beta$, where $\beta \in \{+\alpha, -\alpha, \alpha_H, -\alpha_H \}$, and the corresponding projectors $\hat{P}_\beta = \sum_{j=0}^3 (C^{-1})_{\beta,j} \sigma_j$. The classical transition matrix between the metastable states is given by $\hat{V} = \hat{C}\Lambda\hat{C}^{-1}$, where $\Lambda = \mathrm{diag}(\lambda_0,\dots,\lambda_4)$. This matrix describes the evolution of the jumps within the metastable manifold. To account for the correction, we consider the total evolution as $\mathcal{R}\circ \lv$. This can be evaluated as $r_{\gamma \to \beta} = \tr \hat{P}_\beta \cR(\rho_\gamma)$, which determines the probability that the lobe $\gamma$ is found in $\beta$ after the correction. Thus, the transition matrix with correction is given by
\begin{equation}
    [\hat{V}]_{\alpha \to \beta} = \sum_{\gamma} \insqr{ \hat{C}^{-1} \Lambda \hat{C} }_{\gamma, \alpha} r_{\gamma \to \beta}.
\end{equation}
Upon diagonalization of the symmetric matrix $\hat{V}$, the eigenvector corresponding to the transitions between $+\alpha$ and $-\alpha$ can be identified. The associated eigenvalue $\lambda_{\rm error}$ determines the total bit-flip time of the lobes, such that $\tau_{\rm error} = -1 / \lambda_{\rm error}$. 

\paragraph{Fitness evaluation} The successful execution of all previous steps yields the desired transition rate, conditioned on the existence of the chiral eigenoperator. To prevent the algorithm from failing in the other cases, we use several checks to stop the process as soon as possible and save computational resources. For instance, before diagonalising the Liouvillian, we check whether the mean field solution predicts the existence of the four metastable states. Follow-up checks deal with the existence of metastability or the detection of the chiral eigenoperator. At each halting step, we also adapt the fitness function to guide the algorithm towards a regime where chirality might be present. A summary of these conditions and the corresponding fitness function can be seen in \cref{tab:halting_reasons}.

\begin{table*}
    \centering
    \caption{Minimization function used if the algorithm is not able to calculate the recovery time.}
    \begin{tabular}{l@{\hspace{2em}}|@{\hspace{1em}}c}
    \toprule
        Halt reason & Return value \\ \midrule
        Number of mean field solutions $n$ is smaller than 5 & $\num{1e18}(1 + 5 - n)$ \\
        Mean field mean photon number is different from target & $\num{1e16}(1 + |\abs{\alpha}^2 - \ev{\opn}_{\rm target}|)$ \hspace{1cm} \\
        Chiral eigenoperator not found & $\num{1e14}(1 + \abs{\Re(\lambda_1^{(-1)})})$ \\
        No metastability & $\num{1e13}(1 + \abs{\Re(\lambda_4) / \Re(\lambda_5)})$ \\
        Failed to extract all the metastable states & $\num{1e12}(1 + \abs{\Re(\lambda_{\rm chiral})})$ \\
        Classical $\alpha\to-\alpha$ transition is not found & $\num{1e9}(1 + [\hat{V}]_{\alpha \to -\alpha})$ \\ \bottomrule 
    \end{tabular}
    \label{tab:halting_reasons}
\end{table*}

Finally, for those parameters where the algorithm is able to evaluate $\tau_{\rm error}$, we evaluate the following quantity
\begin{equation}\label{eq:fitness_func_chiral}
    f(\vec{x}) = -\hmt_{\rm gap} \inpar{\tau_{\rm error} - \tau_{\rm leak}}
\end{equation}
where the first term corresponds to the Hamiltonian gap $\hmt_{\rm gap} = E_1 - E_0 > 0$ and the last term is the difference between the bit-flip time ($\alpha \leftrightarrow -\alpha$) with correction and without correction. The latter can also be evaluated from the semiclassical transition matrix $\hat{V}$ as explained above. Defining the minimizing function in this way implies that the larger the Hamiltonian gap, the smaller will it be, as desired. Similarly, this function allows to find those parameters where the chiral correction is more favourable.

\subsection{Results}
We will perform the minimization for $\kappa_1 / 2\pi = \SI{10}{kHz}$, $\kappa_\phi/\kappa_1 \in \{ 0.1, 1, 10 \}$ and $\abs{\alpha}^2 = 2,\dots,15$. We also compare to the results with best known encoding so far, the critical cat. For this, we repeat the simulations without $E_J / E_C$ and replace \cref{eq:fitness_func_chiral} with the fitness function that is just $-\Re(\lambda_{0}^{(-1)})/\hmt_{\rm gap}$ (as in \cref{fig:Kerr_vs_dissipative}).

\begin{figure}
    \centering
    \includegraphics[width=\linewidth]{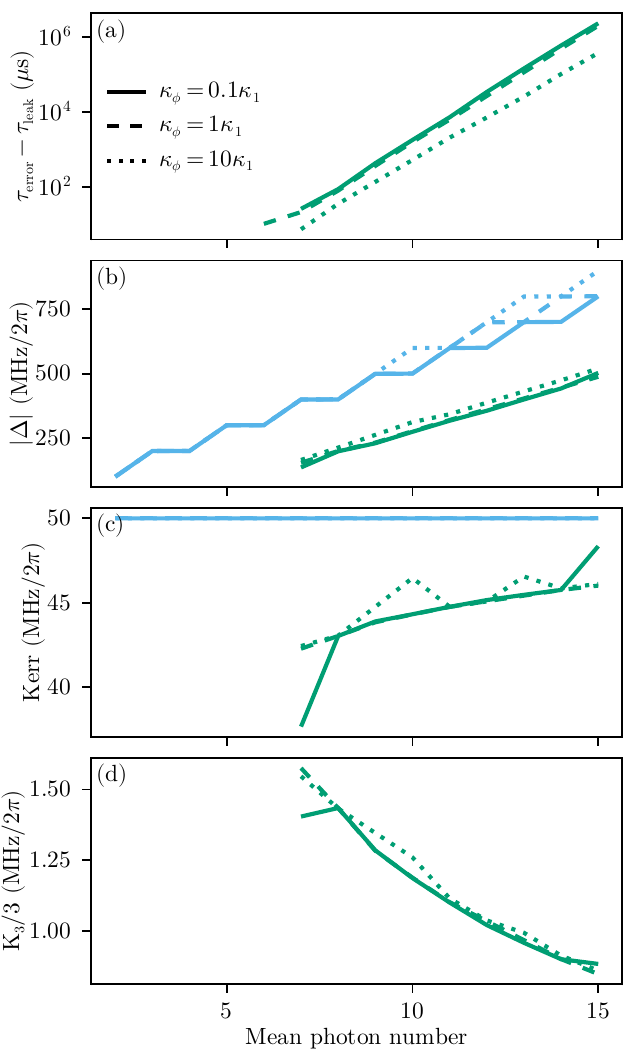}
    \caption{(a) Difference between the bit flip error time with $\tau_{\rm error}$ and without recovery $\tau_{\rm leak}$ as a function of the mean photon number. (b-d) Optimal parameters necessary to minimise the ratio between error rate and Hamiltonian gap of \cref{fig:scale_ratio_photon_number}: (b) detuning and (c) Kerr ($-K_2 / 2$) and (d) first high order correction ($K_3 / 3$).}
    \label{fig:optimal_params}
\end{figure}

The results are presented in \cref{fig:scale_ratio_photon_number} and the optimal parameters in \cref{fig:optimal_params}. As can be seen in panel (a), the chiral correction allows to improve the bit flip error time up to six orders of magnitude with respect to the bit-flip time in uncorrected nonlinear cats. In the other panels (b-d), we show the optimal detuning, Kerr and $K_3$, respectively. Panel (b) confirms the expected trend: larger photon number demands larger detuning, though the chiral cat always operates at a lower optimal detuning than the critical cat. Meanwhile, panel (c) shows the Kerr nonlinearity remains almost constant for all values of $|\alpha|^{2}$, indicating the possibility of operating a single device in the chiral regime by simply detuning and increasing photon drive. We note that the critical cat uses all available nonlinearity for any $\abs{\alpha}^2$, i.e., it hits the upper bound set for this parameter. Increasing the upper bounds of all parameters to $K_2^{\rm max}/2\pi = \SI{50}{MHz}$, $\Delta^{\rm max}/2\pi = \SI{1000}{MHz}$ and $\epsilon_s^{\rm max}/2\pi = \SI{300}{MHz}$, still leads to a saturation of $K_2$ at very large mean photon number. For intermediate values, instead, it converges to a very large $K_2$ or the order of hundreds of MHz and the other parameters, $\Delta$ and $\epsilon_s$, in the order of GHz. These results are unrealistic for state-of-the-art devices. Finally, the first high-order correction $K_3$ is shown in panel (c). Surprisingly, it increases with the mean photon number. However, it remains of the order of a few MHz, which is compatible with a transmon-like nonlinearity. Importantly, using the chiral cat encoding, we can increase the protection against bit flips with feasible experimental parameters.

\section{Quantum trajectories in the displaced Fock basis}
\label{App:Quantum_trajectories}

The quantum trajectories method, also known as the Monte Carlo wave function method, is a stochastic description of open quantum systems \cite{Dalibard1992Wave_function, Dum1992Monte, Molmer1993Monte, Carmichael1993An}. This formalism uses a stochastic Schr\"odinger equation, and the resulting wave function $\ket{\psi(t)}$ is known as a quantum trajectory. Each trajectory evolves continuously by $d\ket{\psi(t)}/dt = - i \hat{H}_{\text{eff}} \ket{\psi(t)}$, where the non-Hermitian effective Hamiltonian is defined by
\begin{equation}
\hat{H}_{\text{eff}} = \hat{H} - \frac{i}{2}\sum_k \hat{L}_k^\dagger \hat{L}_k.
\end{equation}
Here, $\hat{H}$ is the system Hamiltonian defined in Eq.~\eqref{Eq:Hamiltonian}, while $\hat{L}_k$ are jump operators representing the interaction with the environment, which in our case are
the photon loss $\hat{L}_1 = \sqrt{\kappa_1} \hat{a}$, two-photon dissipation $\hat{L}_2 = \sqrt{\kappa_{2}} \hat{a}^2$, and dephasing  $\hat{L}_3 = \sqrt{\kappa_{\phi}} \hat{a}^\dagger \hat{a}$. 
Evolution under $\hat{H}_{\text{eff}}$ is interrupted by quantum jumps, acting as $\hat{L}_k \ket{\psi(t)}$, occurring in an infinitesimal time $dt$ by jump probabilities $p_k = \expval{\hat{L}_k^\dagger \hat{L}_k}{\psi(t)} dt$. Thus, each stochastic trajectory corresponds to a single realization of measurement outcomes of the environment of an ideal measurement instrument.

In an undriven system, quantum jumps corresponding to both one- and two-photon loss progressively collapse the superposition states towards the vacuum. Dephasing, instead, reduces coherence between components without energy loss. 
This is not the case in driven cat states. 
To understand the effect of each term, we resort to the shifted-Fock basis introduced in \cite{chamberland2022} decomposing the annihilation operator into a qubit and gauge sectors, and defined by:
\begin{equation}
    \hat{a}=\hat{Z} \otimes\left(\hat{a}^{\prime}+\alpha\right).
\end{equation}
$\hat{Z}$ is a $2 \times 2$ Pauli $Z$ operator, acting on the qubit sector. $\hat{a}^{\prime}$ is a bosonic annihilation operator of the gauge sector.
$\alpha$ is the amplitude of the coherent state.
The coherent state is then $\ket{\pm \alpha} = \ket{\pm_z, 0}$, so that the cat qubit is the ground state manifold of $\hat{a}'$. For simplicity, we will assume $\alpha$ to be real.
The non-Hermitian Hamiltonian then reads
\begin{equation}
\begin{split}
 \hat{H}_{\rm eff} &= \hat{H} - i \frac{\kappa}{2}\hat{a}^\dagger \hat{a} - i \frac{\kappa_2}{2}\hat{a}^{\dagger\, 2}\hat{a}^2  - i \frac{\kappa_\phi}{2}  (\hat{a}^\dagger \hat{a} )^2 \\ &= 
 \hat{H} - i \frac{\kappa + \kappa_\phi}{2} \hat{a}^\dagger \hat{a} - i \frac{\kappa_2 + \kappa_\phi}{2}\hat{a}^{\dagger\, 2}\hat{a}^2 \\
 &\simeq  - 4 W \alpha^2 \hat{I} \otimes \hat{a}^{\prime \dagger} \hat{a}^{\prime} -  2 W \alpha \hat{I} \otimes\left(\hat{a}^{\prime \dagger 2} \hat{a}^{\prime}+\hat{a}^{\prime \dagger} \hat{a}^{\prime 2}\right) \\ & \quad - c  \alpha \hat{I} \otimes (\hat{a}^{\prime \dagger} + \hat{a}^{\prime})
 ,
 \end{split}
\end{equation}
where $\hat{I}$ is the identity and
\begin{equation}
W = K_2 + i \frac{\kappa_2 + \kappa_\phi}{2}, \quad c = \Delta  + i \frac{\kappa_1  + \kappa_{\phi}}{2}.
\end{equation}
At the same time, the jump operators read
\begin{equation}
\begin{split}
    &\hat{L}_1 = \sqrt{\kappa_1} \hat{a} = \sqrt{\kappa_1} Z \otimes \left(\hat{a}^{\prime}+\alpha\right) \\
    &\hat{L}_2 = \sqrt{\kappa_2} \hat{a}^2 \simeq \sqrt{\kappa_2} \alpha \hat{I}  \otimes \hat{a}^{\prime} \\
    &\hat{L}_3 = \sqrt{\kappa_{\phi}} \hat{a}^\dagger \hat{a}  \simeq \sqrt{\kappa_{\phi}} \alpha \hat{I}  \otimes (\hat{a}^{\prime} + \hat{a}^{\prime \dagger} ).
\end{split}
\end{equation}
Notice that here we neglected $K_3$, as we make the hypothesis confined around the low-excitation code manifold, since we are interested in describing how loss events deteriorate the cat state performance.

These formulas are very instructive.
First, in the non-Hermitian Hamiltonian, the only term ``heating'' up the system is $c = \Delta  + i (\kappa_1  + \kappa_{\phi})/{2}$, exciting the system out of its ground states $\ket{\pm_z,0}$.
As for the quantum jump operators $\hat{L}_1$ and $\hat{L}_2$ associated with $\kappa_1$ and $\kappa_2$, they do not induce any bit flip, but rather cool down the system and ``clean'' the excitation effects induced by the spurious terms $\Delta$, $\kappa_1$ and $\kappa_\phi$.
We conclude that bit-flip errors are due to the lack of quantum jumps induced by one- and two-photon loss, and not because of them.
Instead, dephasing can be interpreted as an effective thermal-like effect, whose strength increases with the number of photons in the system.
Indeed, 
\begin{equation}
    \sqrt{\kappa_{\phi}} \hat{a}^\dagger \hat{a} \ket{\pm\alpha}  \simeq \sqrt{\kappa_{\phi}} \alpha \hat{I}  \otimes  \hat{a}^{\prime \dagger} \ket{\pm_z, 0}.
\end{equation}
This picture highlights the detrimental effect of dephasing.

\bibliography{references}

\end{document}